\documentclass[a4paper,11pt]{article}
\pdfoutput=1 

\usepackage{jheppub} 
\usepackage{CJKutf8}
\usepackage[T1]{fontenc} 

\title{\boldmath Hayward black hole heat engine efficiency in anti-de Sitter space}

\author[a]{Sen Guo,}
\author[b]{Ya-Ling Huang,}
\author[c]{Ke-Jian He,}
\author[d]{Guo-Ping Li}

\affiliation[a]{Guangxi Key Laboratory for Relativistic Astrophysics, School of Physical Science and Technology, Guangxi University, Nanning 530004, People's Republic of China}
\affiliation[b]{School of Physical Science and Technology, SouthWest JiaoTong University, chengdu 610031, People's Republic of China}
\affiliation[c]{College of Physics, ChongQing University, Chongqing 401331, People's Republic of China}
\affiliation[d]{Physics and Space Science College, China West Normal University, Nanchong 637000, People's Republic of China}

\emailAdd{sguophys@126.com}
\emailAdd{katrina996@163.com}
\emailAdd{kjhe94@163.com}
\emailAdd{gpliphys@yeah.net}

\abstract{In this paper, we attempt to further study the heat engine efficiency for the regular black hole (BH) with an anti-de Sitter (AdS) background where the working material is the Hayward-AdS (HAdS) BH. In the extended phase space, we investigate the heat engine efficiency of the HAdS BH by defining the cosmological constant as the thermodynamic pressure $P$ and deriving the mechanical work from the $PdV$ terms. Then, we obtain the relation between the efficiency and the entropy/pressure and plot these function figures. Meanwhile, we compare the relation between the HAdS BH with that of the Bardeen-AdS (BAdS) BH, which is found that the efficiency of the HAdS BH increase with increase the magnetic charge $q$ in contrast to that of the BAdS BH decrease with increase the magnetic charge $q$. We found that the HAdS BH is more efficient than the BAdS BH, and guess that is related to the BH structure.
}

\begin{document}
\maketitle
\flushbottom

\section{Introduction}
\label{sec:intro}
As we all know, the study of the BH singularity has a difficult problem in general relativity. Since singularity cannot be measured by the coordinate transformation, the existence of singularity in the center of BH implies the collapse of the general relativity. Although the theory of quantum gravity is not perfect enough to lead us to solve the classical BH singularity problem. But scientists have built a model of BH solution without singularity in the center, in which the metric and curvature invariants are regular everywhere. Bardeen first proposed BH solution in $1968$ \cite{1}. Ay\'{o}n-Beato $et.al$ pointed out that the physical source of the regular BHs may be the nonlinear electrodynamics \cite{2}, which leads to the extensive study of various regular BH models \cite{3,4,5,6,7}. Based on Bardeen's idea, Hayward proposed a static spherically symmetric BH \cite{5}, the Hayward BH supported by finite density and pressures in which it goes off rapidly at a distance increases and treated as a cosmological constant at radius fall off very near to the origin. It is curvature invariants being everywhere finite and satisfying the weak energy condition. Subsequently, the geodesic equation of the Hayward BH has been investigated \cite{8}. The implication of the rotating and modified Hayward BH is discussed in \cite{9,10}.

BH as the thermodynamic system has many interesting consequences.  It connects the quantum aspects of space-time geometry with classical thermodynamic theory and serves as a bridge between general relativity, thermodynamics, and quantum mechanics. The origin of this research can be traced back to the pioneering work of the Hawking and Bekenstein, who first proposed the Hawking temperature and Bekenstein-Hawking entropy \cite{11,12,13}. Subsequently, Hawking and Page have first investigated the Schwarzschild-AdS BH and the thermal AdS space first-order phase transition \cite{14}. Chamblin et.al found the RN-AdS BH is similar to the van der Waals (vdW) liquid/gas phase transition \cite{15}. In the AdS space, the negative cosmological constant is treated as thermodynamic pressure \cite{16}, it is can be written as
\begin{equation}
P=-\frac{\Lambda}{8\pi},
\label{1}
\end{equation}
and the pressure conjugate quantity is considered to be thermodynamic volume \cite{16,17}
\begin{equation}
V={\Big(\frac{\partial M}{\partial P}\Big)}_{S,Q,J}.
\label{2}
\end{equation}
The $PdV$ term appears in the BH thermodynamics first law \cite{18}. Inspired by the above research, many kinds of research have shown that the AdS BH phase transition is similar to that of the vdW system in the extended phase space. The BH shares the same $P-V$ diagram and critical exponents with the vdW system \cite{19,20,21,22,23,24,25,26,27,28,29,30,31,32}.

Apart from the phase transition and critical phenomena, the recent developments in BH thermodynamics are the Joule-Thomson expansion \cite{33,34,35,36,37,38} and the weak cosmic censorship conjecture \cite{39,40,41}, $etc.$ On the other hand, the Penrose process can extract energy from the rotating BHs in both asymptotically AdS and flat space-time. However, it is interesting that the holographic heat engine can extract mechanical work via from the $PdV$ term, which is the opposite of the Penrose process. A heat engine is defined by a heat cycle. It is a closed path in the $P-V$ diagram, which is allow the input heat $Q_{H}$ and exhaust heat $Q_{C}$. The total mechanical work done is $W=Q_{H}-Q_{C}$ by the thermodynamic first law. The efficiency of the heat engine is defined by $\eta={W}/{Q_{H}}=1-{Q_{H}}/{Q_{C}}$, which is clear that the efficiency of the BH heat engine depends crucially on both heat absorbed and work done provided by the BH. Johnson built the first holographic BH heat engine, and obtained the conversion efficiency \cite{42,43}. Subsequently, the concept of holographic heat engine has been extended to the various BH backgrounds, such as the static and dynamic BHs \cite{44}, the polytropic BH \cite{45}, the Born-Infeld-AdS BH \cite{46}, $etc$ \cite{47,48,49,50,51,52,53,54,55,56}.

Recently, Fan studied the critical phenomena and first-order small and large BH phase transition of the HAdS BH \cite{57}. More recently, Mehdipour et.al investigated the thermodynamics and phase transition of Hayward solutions \cite{58}. In this paper, we will investigate the thermal efficiency of the HAdS BH, and shows the difference between the HAdS BH and BAdS BH of the heat engine efficiency.

The remainder of this paper is organized as follows. In Sec.\ref{sec1}, we review the thermodynamic properties of the HAdS BH in the extended phase space. In Sec.\ref{sec2}, we investigate the heat engine efficiency of this BH. In Sec.\ref{sec3}, we compare the heat engine efficiency of the HAdS BH with the BAdS BH. Sec.\ref{sec4}, it ends up with some conclusions.

\section{The HAdS BH thermodynamic}
\label{sec1}

The HAdS BH metric has the form \cite{59}.
\begin{equation}
ds^{2}=-f(r)dt^{2}+\frac{dr^{2}}{f(r)}+r^{2}d\Omega^{2},
\label{3}
\end{equation}
where
\begin{equation}
f(r)=1+\frac{r^{2}}{l^{2}}-\frac{2 M r^{2}}{r^{3}+q^{3}},
\label{4}
\end{equation}
in which $l^{2}={3}/{8 \pi P}$, $q$ is the integration constant with respect to magnetic charge, $M$ is viewed as the BH mass. The event horizon of the BH is articulated by $f(r_{h})=0$, which gives the BH mass
\begin{equation}
M=\frac{(3+8 P \pi r_{h}^{2})(q^{3}+r_{h}^{3})}{6r_{h}^{2}}.
\label{5}
\end{equation}
The entropy of the BH is $S=\pi r_{h}^2$, the BH temperature is
\begin{equation}
T=\frac{-2 \pi^{{3}/{2}}q^{3}+S^{{3}/{2}}(1+8PS)}{4\sqrt{\pi}S^{2}}.
\label{6}
\end{equation}
We obtain the equation of state for the HAdS BH
\begin{equation}
P=\frac{T}{2 r_{h}}+\frac{1}{8 \pi {r_{h}}^{2}}+\frac{q^{3}}{4 \pi {r_{h}}^{5}}.
\label{7}
\end{equation}
The critical points are obtained from the conditions ${\partial P}/{\partial \nu}=0={\partial^2 P}/{\partial \nu^2}$, where the specific volume $\nu$ is twice as much the outer horizon radius $r_{+}$($\nu=2r_{+}$),
\begin{eqnarray}
\nu_{c}=2^{5/3}\times5^{1/3}q,~~P_{c}=\frac{3}{80\times2^{1/3}\times5^{2/3}\pi q^{2}},~~T_{c}=\frac{3}{2^{11/3}\times5^{1/3}\pi q}.
\label{8}
\end{eqnarray}
Other thermodynamic properties can be obtained by using above relations, such as heat capacities at constant pressure and constant volume, i.e.
\begin{equation}
C_{P}=T\Big(\frac{\partial S}{\partial T}\Big)_{P,q}=\frac{2S(-2 \pi^{{3}/{2}}q^{3}+S^{{3}/{2}}(1+8P S))}{8 \pi^{{3}/{2}}q^{3}-\sqrt{\pi}S+8 P S^{{5}/{2}}},
\label{9}
\end{equation}
and
\begin{equation}
C_{V}={T\Big(\frac{\partial S}{\partial T}\Big)}_{V,Q}=0.
\label{10}
\end{equation}

\begin{figure}[h]
\centering 
\includegraphics[width=.45\textwidth]{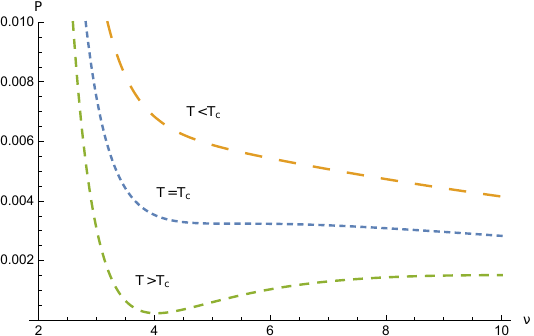}
\includegraphics[width=.45\textwidth]{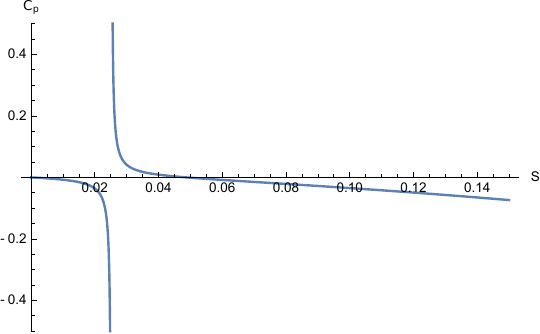}
\caption{\label{fig1} $P-\nu$ diagram for the HAdS BH with $q=1$ for different values of temperature in left figure. $C_{P}-S$ diagram for HAdS BH with $q=1$ in right figure.}
\end{figure}
We plot the $P-\nu$ isotherm and $C_{P}-S$ curves using the equations (\ref{7}) and (\ref{8}) as shown in figure (\ref{fig1}). The critical isotherm ($T=T_{c}$) has an inflection point. Above the critical isotherm ($T>T_{c}$), there is no inflection point, and pressure is a monotonically decreasing function of radius. Below the critical isotherm ($T<T_{c}$), there are two extrema(one maximum and one minimum), and the slope region is positive ($T<T_{c}$,$\partial P / \partial r >0 $), which thermodynamic instability occurs. On the other hand, $C_{P}$ shows a negative and positive trend, the specific heat for small BH region and large BH region means that those BHs are thermodynamically stable. The intermediate BH region represents unstable system. The actual phase transition takes place between small BH and large BH.


\section{HAdS BH as a Heat Engine}
\label{sec2}

In this section, we attempt to calculate the heat engine efficiency of the HAdS BH. As we all know, the heat engine is constructed as a closed path in the $P-V$ plane, which the heat absorbed is defined as $Q_{H}$ and the heat discharged is defined as $Q_{C}$ (Fig.2). According to the thermodynamic first law, the total mechanical work is $W=Q_{H}-Q_{C}$, the efficiency of the heat engine is determined by \cite{46}
\begin{equation}
\eta=\frac{W}{Q_{H}}.
\label{11}
\end{equation}

In the classical thermodynamics, there are several thermodynamical cycles that are used in heat engine: the Carnot cycle, the Otto cycle and the Brayton cycle. The Carnot cycle is made of two pairs of isothermal and adiabatic processes and has the highest efficiency, which the efficiency is given by
\begin{equation}
\eta_{c}=1-\frac{Q_{C}}{Q_{H}}=1-\frac{T_{C}}{T_{H}},
\label{12}
\end{equation}
where $T_{C}$ and $T_{H}$ are the lower and higher temperatures of the process figure 2.
\begin{figure}[h]
\centering 
\includegraphics[width=.45\textwidth]{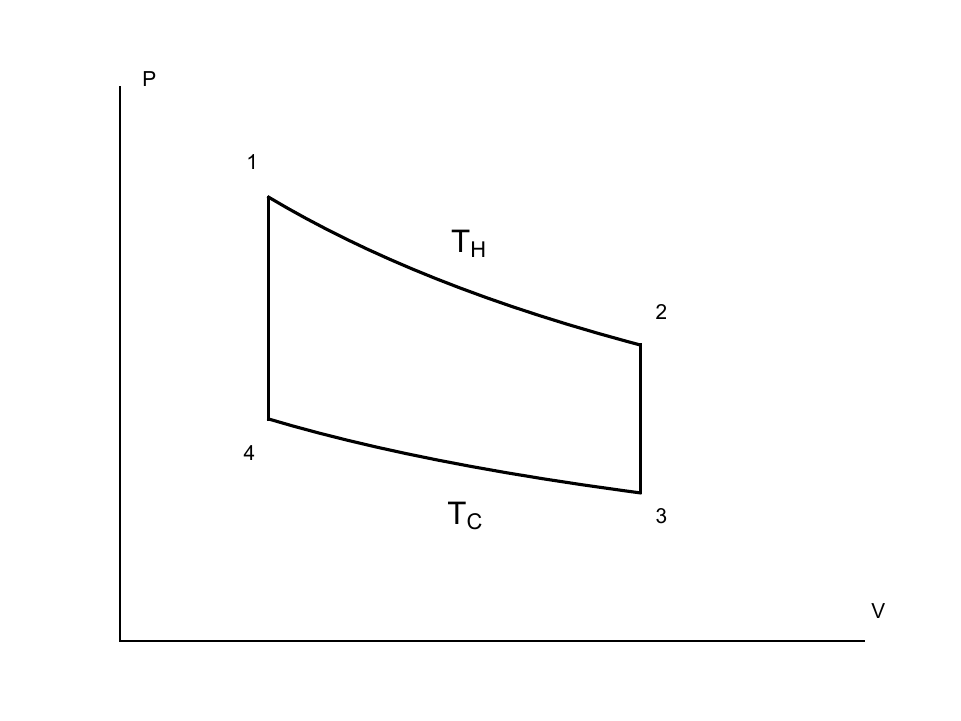}
\includegraphics[width=.45\textwidth]{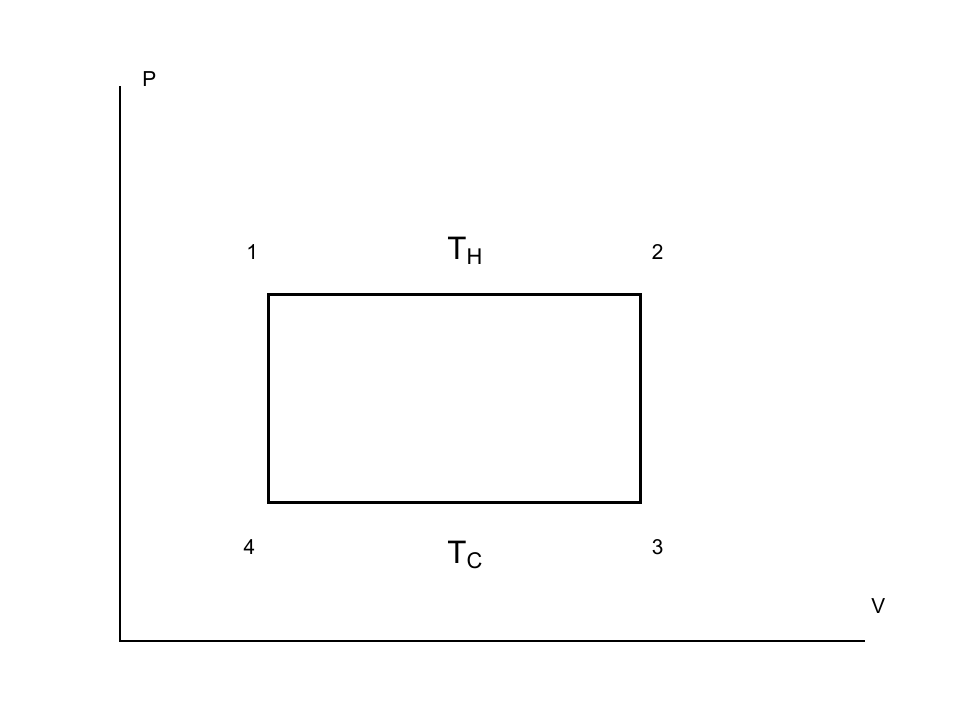}
\caption{\label{fig2} The left figure is the Carnot engine cycle. The right figure is the HAdS BH heat engine.}
\end{figure}

We are attempting to construct a simple heat engine on the background of the BH. The right half of the figure (\ref{fig2}), this cycle is made of two pairs of the isobars and adiabats. For simplicity, we consider a rectangular cycle ($1 \rightarrow 2 \rightarrow 3 \rightarrow 4 \rightarrow 1$). And, we choose the cycle to occur in the stable region of small BH and large BH. Without considering the phase transition overlapping region in the middle, so as to define the heat engine and improve the efficiency. In the process of isothermal expansion and compression, the heat absorbed $Q_{H}$ and heat discharged $Q_{C}$ can be written as
\begin{equation}
Q_{H}=T_{H}\Delta S_{1\rightarrow 2}=T_{H}{\Big(\frac{3}{4 \pi}\Big)}^{{2}/{3}}\pi (V_{2}^{{2}/{3}}-V_{1}^{{2}/{3}}),
\label{13}
\end{equation}
\begin{equation}
Q_{C}=T_{C}\Delta S_{3\rightarrow 4}=T_{C}{\Big(\frac{3}{4 \pi}\Big)}^{{2}/{3}}\pi (V_{3}^{{2}/{3}}-V_{4}^{{2}/{3}}).
\label{14}
\end{equation}
where the $T_{H}$ and $T_{C}$ correspond to the stable temperature of small BH and large BH respectively, the $V_{1}$ and $V_{2}$ correspond to the stable volume of small BH and large BH.

If we choose isochores to connect those isotherms, the $V_{2}=V_{3}$ and $V_{1}=V_{4}$ can be obtained. The equations (\ref{13}) and (\ref{14}) are lead
\begin{equation}
\eta=1-\frac{Q_{C}}{Q_{H}}=1-\frac{T_{C}}{T_{H}}.
\label{15}
\end{equation}
It is same as the efficiency of the Carnot engine. As figure 2, the area of a rectangle can be defined as the work done
\begin{eqnarray}
W=&&\oint PdV=P_{1}(V_{2}-V_{1})+P_{4}(V_{4}-V_{3}) \nonumber\\
=&&\frac{4}{3\sqrt{\pi}}(P_{1}-P_{4})(S_{2}^{{3}/{2}}-S_{1}^{{3}/{2}}).
\label{16}
\end{eqnarray}
According to Eq. (\ref{14}), one can get the processes $1 \rightarrow 4$ and $2 \rightarrow 3$ have not heat exchange. Hence, we only need calculate heat absorbed $Q_{H}$ in the process $1 \rightarrow 2$ \cite{60}, i.e.
\begin{eqnarray}
Q_{H}=&&\int^{T_{2}}_{T_{1}}C_{P}(P_{1},T)dT=\int^{r_{2}}_{r_{1}} \Big(\frac{\partial M}{\partial r}\Big) dr \\ \nonumber
=&&\frac{1}{6\sqrt{\pi}}\Big[\frac{3 {\pi}^{{3}/{2}}q^{3}+S_{2}^{{3}/{2}}(3+8P_{1}S_{2})}{S_{2}}-\frac{3 {\pi}^{{3}/{2}}q^{3}+S_{1}^{{3}/{2}}(3+8P_{1}S_{1})}{S_{1}}\Big].
\label{17}
\end{eqnarray}
Therefore, the heat engine efficiency of the HAdS BH can be written as
\begin{eqnarray}
\eta=\frac{W}{Q_{H}}=\frac{8S_{1}S_{2}(P_{1}-P_{4})(S_{2}^{{3}/{2}}-S_{1}^{{3}/{2}})}{{S_{1}(3{\pi}^{{3}/{2}}q^{3}+S_{2}^{{3}/{2}}(3+8P_{1}S_{2}))}-{S_{2}({3 {\pi}^{{3}/{2}}q^{3}+S_{1}^{{3}/{2}}(3+8P_{1}S_{1}))}}}.
\label{18}
\end{eqnarray}
\par
According to equation (\ref{17}), we compare the heat engine efficiency of the HAdS BH and Carnot engine ($\eta_{C}$). Taking the the higher temperature $T_{H}$ ($T_{2}$) and lower temperature $T_{C}$ ($T_{4}$), the Carnot heat engine efficiency is obtained, i.e.
\begin{equation}
\eta_{c}=1-\frac{T_{4}(P_{4},S_{1})}{T_{2}(P_{1},S_{2})}=1-\frac{S_{2}^{2}(-2 \pi^{{3}/{2}}q^{3}+S_{1}^{{3}/{2}}(1+8P_{4}S_{1}))}{S_{1}^{2}(-2 \pi^{{3}/{2}}q^{3}+S_{2}^{{3}/{2}}(1+8P_{1}S_{2}))}.
\label{19}
\end{equation}

We plot the $\eta-S_{2}$ curves and $K-S_{2}$ curves using the equations (\ref{18}) and (\ref{19}) as shown in figure 3, where the $K\equiv \eta/\eta_{C}$. The heat engine efficiency is a monotonously increasing/decreasing function with the growth of the entropy $S_{2}$. The heat engine efficiency increases with the increase of $S_{2}$ when the $q$ is small. While the efficiency of the heat engine decreases first and then increases with the increase of $S_{2}$ when the $q$ is large. The heat engine efficiency tends to reach a saturation value with the increase of entropy. The $K-S_{2}$ curves also shows similar characteristics.

\begin{figure}[h]
\centering 
\includegraphics[width=.45\textwidth]{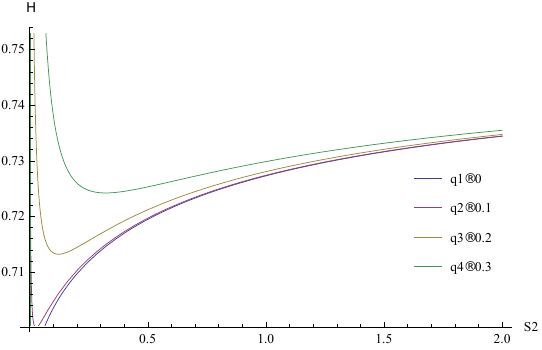}
\includegraphics[width=.45\textwidth]{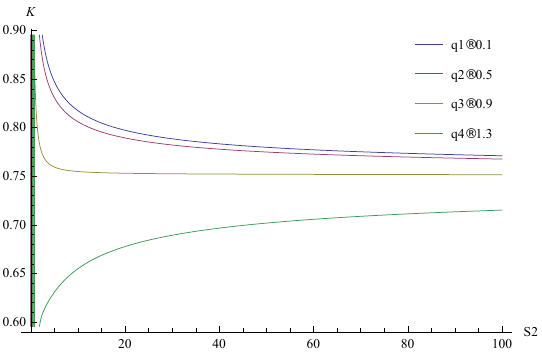}
\caption{\label{fig3} $\eta-S_{2}$ diagram for the HAdS BH for different values of $q$ in left figure. The right figure shows $K-S_{2}$ diagram of the HAdS BH with different values of $q$. We take $P_{1}=4, P_{4}=1, S_{1}=1$.}
\end{figure}

We also investigate the relationship between the heat engine efficiency and the pressure $P_{1}$ as shown in figure 4. It is shows that the heat engine efficiency increase with increase pressure and tends to maximum. And, the heat engine efficiency of the HAdS BH increase gradually with the increase of $q$.
\begin{figure}[h]
\centering 
\includegraphics[width=.45\textwidth]{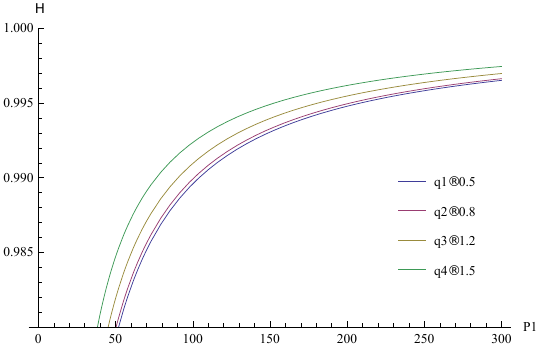}
\includegraphics[width=.45\textwidth]{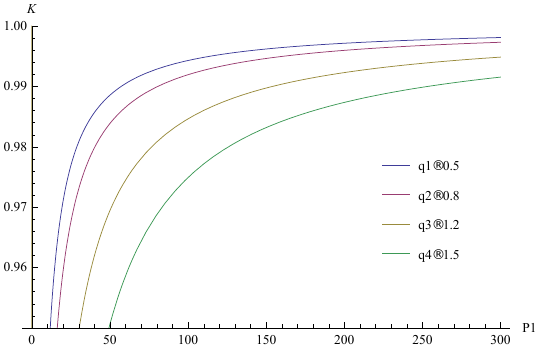}
\caption{\label{fig4} $\eta-P_{1}$ diagram for the HAdS BH for different values of $q$ in left figure. The right figure shows $K-P_{1}$ diagram of the HAdS BH with different values of $q$. We take $S_{2}=4, P_{4}=1, S_{1}=1$.}
\end{figure}

\section{The HAdS BH heat engine and BAdS BH heat engine}
\label{sec3}
In this section, we will compare the HAdS BH and BAdS BH in terms of heat engine efficiency. In \cite{55}, the BAdS BH is defined as the heat engine. The BAdS BH heat engine efficiency is obtained, and the functional image between the efficiency $\eta$ and entropy/pressure is plotted. The BAdS BH heat engine efficiency is given by \cite{55}
\begin{equation}
{\eta}_{B}=\frac{W}{Q_{H}}=\frac{4(P_{1}-P_{4})(S_{2}^{{3}/{2}}-S_{1}^{{3}/{2}})}{3\sqrt{\pi}\Big(\frac{\pi (q^{2}+{S_{2}}/{\pi})^{{3}/{2}}(8P_{1}S_{2}+3)}{6S_{2}}-\frac{\pi (q^{2}+{S_{1}}/{\pi})^{{3}/{2}}(8P_{1}S_{1}+3)}{6S_{1}}\Big)}.
\label{20}
\end{equation}
Taking the $P_{1}=4, P_{4}=1, S_{1}=1, S_{2}=4, q=1$, the heat engine efficiency of the HAdS BH is higher than that of the BAdS BH according to equations (\ref{18}) and (\ref{20}). And, we can get the relationship between them, i.e.
\begin{equation}
\frac{{\eta}_{H}}{{\eta}_{B}}=1.53884.
\label{21}
\end{equation}
Moreover, the relation between the BAdS BH heat engine efficiency and entropy $S_{2}$ is plotted in figure 5.
\begin{figure}[h]
\centering 
\includegraphics[width=.45\textwidth]{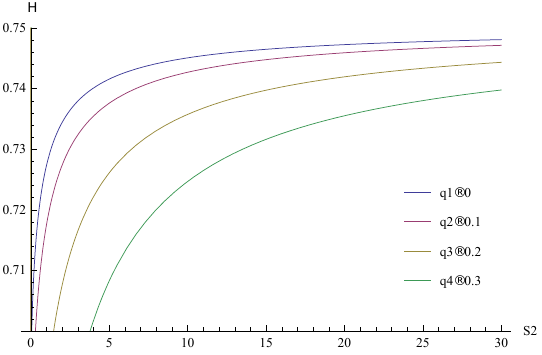}
\includegraphics[width=.45\textwidth]{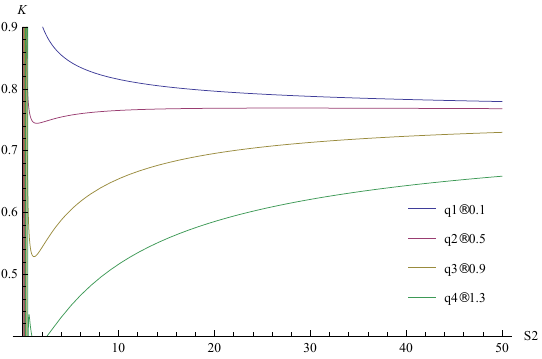}
\caption{\label{fig5} $\eta-S_{2}$ diagram for the BAdS BH for different values of $q$ in left figure. The right figure shows $K-S_{2}$ diagram of the BAdS BH with different values of $q$. We take $P_{1}=4, P_{4}=1, S_{1}=1$.}
\end{figure}

It is shown that the BAdS BH heat engine efficiency monotonously increases with $S_{2}$ for different values of $q$, in which the increase in volume difference between small BH ($V_{1}$) and large BH ($V_{2}$) also increases the efficiency. The change form of $K-S_{2}$ is also related to the size of $q$, this ratio is also saturated when $S_{2}$ reaches a specific value.

The relation between the BAdS BH heat engine efficiency and pressure is shown in figure 6. It is means that the heat engine efficiency increase monotonically with the pressure increase, and finally approaches the maximum efficiency.

\begin{figure}[h]
\centering 
\includegraphics[width=.45\textwidth]{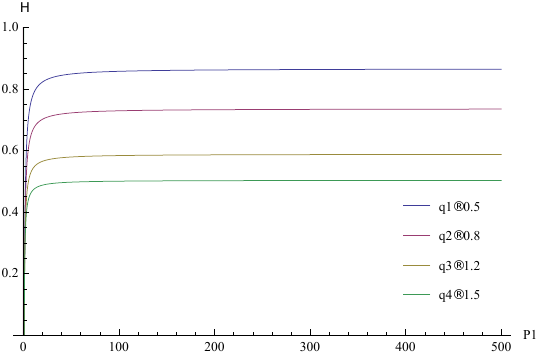}
\includegraphics[width=.45\textwidth]{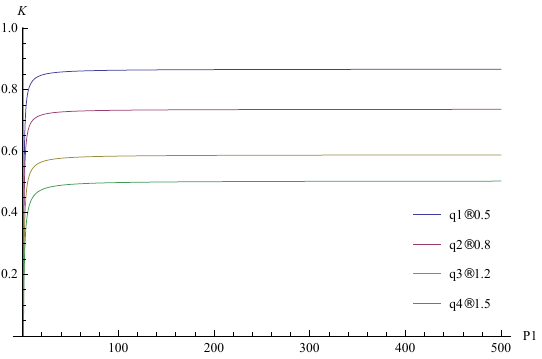}
\caption{\label{fig6} $\eta-P_{1}$ diagram for the HAdS BH for different values of $q$ in left figure. The right figure shows $K-P_{1}$ diagram of the HAdS BH with different values of $q$. We take $S_{2}=4, P_{4}=1, S_{1}=1$.}
\end{figure}

According to figures 3 and 5, we can seen that the BAdS BH heat engine efficiency increases with the increase of the entropy. While this efficiency decreases gradually with the increase of $q$. For the HAdS BH, the heat engine efficiency is a monotonously increasing/decreasing function with the growth of the entropy $S_{2}$. The heat engine efficiency increases with the increase of $S_{2}$ when the $q$ is small. While the efficiency of the heat engine decreases first and then increases with the increase of $S_{2}$ when the $q$ is large. The heat engine efficiency tends to reach a saturation value with the increase of entropy. The figures 4 and 6 shown that the relation between the heat engine efficiency and pressure. The BAdS BH heat engine efficiency increases with the increase pressure, but the heat engine efficiency slows down with the increase $q$. For the HAdS BH, the heat engine efficiency increases gradually with the increase pressure, and tends to reach a stable maximum efficiency. Interestingly, the heat engine efficiency of the HAdS BH increases with the increase $q$ at the same pressure, whereas that of the BAdS BH is the opposite.

\section{Conclusions and discussions}
\label{sec4}

In this paper, we investigate the heat engine efficiency for the HAdS BH as a heat engine. We obtain the relationship between the heat engine efficiency and entropy/pressure, which is compared with that of the BAdS BH. In the extended phase space, the equation of state of the HAdS BH is obtained by viewing the cosmological constant as the thermodynamic pressures. The heat engine efficiency of the HAdS BH is determined by the work done and the heat absorbed in the cycle, which order to there are two isotherms and two isochores in the $P-V$ plane. Comparing the Carnot heat engine as a standard heat engine, we obtain the relation between the HAdS BH heat engine efficiency and entropy/pressure, analyzing the engine efficiency with the change of the magnetic charge $q$. We interestingly find that the heat engine efficiency is a monotonously increasing/decreasing function with the growth of the entropy $S_{2}$. The heat engine efficiency increases with the increase of $S_{2}$ when the $q$ is small. While the efficiency of the heat engine decreases first and then increases with the increase of $S_{2}$ when the $q$ is large. The heat engine efficiency tends to reach a saturation value with the increase of entropy.

More importantly, we compare the heat engine efficiency of the HAdS BH with that of the BAdS BH. It is found that the mechanical efficiency of the HAdS BH is higher than that of the BAdS BH. It is reveals that the heat engine efficiency of the HAdS BH increases with the increase of the magnetic charge $q$ in contrast to that of the BAdS BH, which decreased with the increase of the magnetic charge $q$. The heat engine efficiency of the HAdS BH is not only related to the magnetic charge $q$ but also to the range of the entropy in contrast to that of the BAdS BH, which only depended on the magnetic charge $q$.


\acknowledgments

The authors would like to thank the anonymous reviewers for their helpful comments and suggestions, which helped to improve the quality of this paper. This work is supported by the National Natural Science Foundation of China No. 11903025 and the Fundamental Research Funds of China West Normal University (cxcy2018240).



\end{document}